\documentclass[secnumarabic,amssymb, nobibnotes, aip, jmp]{revtex4-1}

\usepackage{graphicx}
\usepackage{dcolumn}
\usepackage{bm}
\usepackage{hyperref}
\usepackage{CJK}

\begin{document}


\title{Analytical solutions for smooth positive-to-negative transition materials}

\author{Yi S. Ding}
\affiliation{Department of Physics, The Hong Kong University of Science and Technology and School of physics, Peking University}

\author{C. T. Chan}
\affiliation{Department of Physics, The Hong Kong University of Science and Technology}



\date{\today}

\begin{abstract}
We obtain analytical solutions for positive-to-negative transition materials with a smooth transition profile ``$\tanh$''. The fields are expressed in terms of hypergeometric functions. The final expressions for reflection and transmission coefficients are obtained in closed forms. The total absorption is nonzero even in the lossless limit in accordance with previous studies. Properties of the total absorption as function of controlling factors are also well studied. As an application of the analytical results, we analyze the effects of smooth transition on perfect imaging and, interestingly, find multiple images of the object.

\end{abstract}

\maketitle

\section{Introduction}
Positive-to-negative transition materials are layered optical materials with gradient optical indexes $\epsilon$ and/or $\mu$ continuously changing from positive values to negative ones. In Ref.\onlinecite{Landau} and \onlinecite{Ginzburg}, it is obtained that near the transition point for $\epsilon$, the fields for oblique incidence present large enhancement with enhanced absorption. Actually the absorption is nonzero even in the lossless limit. In Ref. \onlinecite{Litchinitser2008}, similar effects are obtained when both $\epsilon$ and $\mu$ linearly pass the same transition point, and the fields are expressed in terms of confluent hypergeometric functions near that point.  Also in Ref. \onlinecite{Litchinitser2008}, smooth transition material $\epsilon=\mu=\tanh x$ (tanh-material) is only studied numerically. In Ref. \onlinecite{Dalarsson2009}, the properties for the ``tanh'' profile under normal incidence are studied, but no field enhancement is present for normal incidence. In this study, we obtain analytical solutions for the ``tanh'' material under oblique incidence. The fields are expressed in terms of hypergeometric functions, and the reflection as well as the transmission coefficients are expressed in closed-form analytical formulas. The absorption in the lossless limit is nonzero, as expected. Therefore, this study provides an example of transition materials \cite{Litchinitser2012,Litchinitser2011,Litchinitser2010,Litchinitser2011B} that can be treated fully analytically

As an application of the obtained formulas, we also analyze the effects of smooth transition on perfect imaging. In the sharp transition limit, the material can have perfect imaging effect, i.e., all evanescent waves can be ideally recovered \cite{Pendry2000}. However, for smooth transition, it is not that case. Our analytical results can shed light on the effects of smooth transition. The main feature of the image plane is that it contains multiple images of an object.

\section{Helmholtz equation for steady states}
The permittivity and permeability we are interested in are $\epsilon=\mu=\tanh(\rho x+i\delta)$. The imaginary part is positive for $0<\delta<\pi/2$, and therefore the material is passive under that condition.

The Helmholtz equation for TE waves are [\onlinecite{Litchinitser2008}],
\begin{equation}
\frac{d^{2}E}{dx^{2}}-\frac{1}{\mu}\frac{d\mu}{dx}\frac{dE}{dx}+(a^{2}\mu^{2}-b^{2})E=0.\label{helmholtz}
\end{equation}
where $a=\omega/\rho$, $b=\omega\sin \theta/\rho$ with $\omega$ and $\theta$ being the frequency and angle of the incident wave.

The Eq.\ref{helmholtz} can be transformed into hypergeometric equation after changes of variables, $y=\mu^2$ and $E=(1-y)^\lambda F(y)$ where $\lambda=-\frac{i}{2}\sqrt{a^2-b^2}$. The resulted equation is
\begin{equation}
\frac{d^{2}F}{dy^{2}}-\frac{2\lambda+1}{1-y}\frac{dF}{dy}-\frac{b^{2}}{4y(1-y)}F=0,\label{hypergeom_eq}
\end{equation}
which is a hypergeometric equation with indexes $\alpha=\lambda+ia/2$, $\beta=\lambda-ia/2$ and $\gamma=0$.

\section{Solutions for $x\rightarrow+\infty$, $x\rightarrow -\infty$ and $-\infty<x<+\infty$.}
For $x\rightarrow\pm\infty$, we have $\mu\rightarrow\pm1$ and $y\rightarrow1$. According to Appendix A, the field solution can be expressed as linear superposition of two basic solutions,
\begin{equation}
E_\pm(x)=C_\pm^{(1)}(1-\mu^2)^{\lambda}w_3+C_\pm^{(2)}(1-\mu^2)^{\lambda}w_4,\label{Epm}
\end{equation}
where $w_{3,4}$, as well as $w_{1,2}$ below, are Kummer solutions for hypergeometric equations in Appendix A.

For $-\infty<x<+\infty$, $y$ is near the zero pole. The two basic solutions are,
\begin{equation}
(1-\mu^2)^{\lambda}f \,\,\,\,\mbox{and}\,\,\,\,(1-\mu^2)^{\lambda}w_2.
\end{equation}
where $w_2$ is one of the Kummer solutions near the zero pole [\onlinecite{Nist}] and $f$ is the solution with logarithm since we are working with a degenerate hypergeometric equation with index $\gamma=0$.

Similar to the expression of Neumann functions in terms of Bessel function, the logarithm solution $f$ can be expressed as limit of linear superposition of $w_1$ and $w_2$ of nonzero $\gamma$ index,
\begin{equation}
f=\lim_{\gamma\rightarrow0}\frac{\frac{w_{1}}{\Gamma(\gamma)}-\alpha\beta w_{2}}{\frac{1}{\Gamma(\gamma)}},\label{f}
\end{equation}
which will be proved in the Appendix B. Specifically, according to the  L 'Hospital rule,
\begin{equation}
f=\lim_{\gamma\rightarrow0}\frac{\frac{w'_{1}\Gamma(\gamma)-w_{1}\Gamma'(\gamma)}{\Gamma(\gamma)^{2}}-\alpha\beta w'_{2}}{-\frac{\Gamma'(\gamma)}{\Gamma(\gamma)^{2}}},\label{f_lh}
\end{equation}
where $w'_1=\frac{dw_1}{d\gamma}$, $w'_2=\frac{dw_2}{d\gamma}=- w_{2}\ln y+y^{1-\gamma}\frac{d}{d\gamma} ~_2F_1(\alpha-\gamma+1,\beta-\gamma+1;2-\gamma;y)$ with $_1F_2$ being the hypergeometric function.

However, if we use the principal values of $f(y)$, the field is discontinuous \cite{Ingrey2008} along the real $x$ axis at zero point $x=0$, corresponding to $\mu=\tanh(i\delta)=i\tanh\delta$. Therefore, for $0<x<+\infty$ and $-\infty<x<0$, we should use different superpositions of $w_2$ and $f$ to make sure that the field is continuous.

The discontinuity of the principal value of $f$ is $f(x=0+)-f(x=0-)=2\alpha\beta\pi iw_2(x=0)$ according to Eq.(\ref{f_lh}). Then, if the field for $0<x<+\infty$ is \begin{equation}
E_+(x)=A(1-\mu^2)^{\lambda}f+B(1-\mu^2)^{\lambda}w_2,\label{Ep}
\end{equation}
the field in $-\infty<x<0$ should be
\begin{equation}
E_-(x)=A(1-\mu^2)^{\lambda}f+(B+2\alpha\beta\pi i A)(1-\mu^2)^{\lambda}w_2,\label{Em}
\end{equation}
to ensure the continuity across $x=0$.

\section{relations among $C_{\pm}^{(1,2)}$ and $A$, $B$}
The different field expressions (\ref{Epm}, \ref{Ep},\ref{Em}) should be equal in equivalent regions, which imposes connection relations among the coefficients $C_{\pm}^{(1,2)}$ and $A$, $B$.

If we set $C_+^{(1)}=1$, $C_+^{(2)}=0$ in $x\rightarrow+\infty$ region, according to the Kummer connection formula between $w_3$ and $w_{1,2}$ in Appendix A, as well as Eq. (\ref{f},\ref{Ep}), we have
\begin{eqnarray}
w_{3}&=&\frac{\Gamma(1-\gamma)\Gamma(\alpha+\beta-\gamma+1)}{\Gamma(\alpha-\gamma+1)\Gamma(\beta-\gamma+1)}w_{1}+
\frac{\Gamma(\gamma-1)\Gamma(\alpha+\beta-\gamma+1)}{\Gamma(\alpha)\Gamma(\beta)}w_{2}\nonumber\\
&=&[B(\gamma)-A(\gamma)\alpha\beta\Gamma(\gamma)]w_{2}+A(\gamma) w_{1}.\label{ForAB}
\end{eqnarray}
with $B=\lim_{\gamma\rightarrow0}B(\gamma)$,$A=\lim_{\gamma\rightarrow0}A(\gamma)$. After straightforward calculation, we have
\begin{eqnarray}
A&=&\frac{\Gamma(\alpha+\beta+1)}{\alpha\beta\Gamma(\alpha)\Gamma(\beta)}\nonumber\\
B&=&\frac{1}{\Gamma(\alpha)\Gamma(\beta)}(\frac{1}{\alpha}+\frac{1}{\beta}+\gamma_{E}+\psi(\alpha)+\psi(\beta)-1).
\end{eqnarray}
where $\gamma_{E}=-\psi(1)=0.577215664....$ and $\psi$ is the derivative
of the logarithm of $\Gamma$ function.

Then according to Eq.(\ref{Ep},\ref{Em},\ref{ForAB}), in $x\rightarrow-\infty$
\begin{equation}
E_-(x)=(1-\mu^2)^\lambda w_3+2\alpha\beta\pi i A (1-\mu^2)^\lambda w_2.
\end{equation}

By use of Kummer connection formula between $w_2$ and $w_{3,4}$ in Appendix A, we have
\begin{equation}
E_-(x)=\left[1+\frac{A\alpha\beta(2\pi i)\Gamma(-\alpha-\beta)}{\Gamma(1-\alpha)\Gamma(1-\beta)}\right](1-\mu^2)^\lambda w_{3}+\frac{A\alpha\beta(2\pi i)\Gamma(\alpha+\beta)}{\Gamma(\alpha+1)\Gamma(\beta+1)}(1-\mu^2)^\lambda w_{4},
\end{equation}
which, compared with Eq.(\ref{Em}), gives
\begin{eqnarray}
C_-^{(1)}&=&1+\frac{(2\pi i)\Gamma(-\alpha-\beta)}{\Gamma(1-\alpha)\Gamma(1-\beta)}\frac{\Gamma(\alpha+\beta+1)}{\Gamma(\alpha)\Gamma(\beta)}
=1-i\frac{2\sin\pi\alpha\sin\pi\beta}{\sin\pi(\alpha+\beta)}\nonumber\\
C_-^{(2)}&=&\frac{(2\pi i)\Gamma(\alpha+\beta)}{\Gamma(\alpha+1)\Gamma(\beta+1)}\frac{\Gamma(\alpha+\beta+1)}{\Gamma(\alpha)\Gamma(\beta)}
\end{eqnarray}

\section{reflection and transmission coefficients}\label{SecRT}
Now we have obtained continuous solutions along the real $x$ axis. For $x\rightarrow+\infty$, $E_+=(1-\mu^2)^\lambda w_{3}$; for $x\rightarrow-\infty$, $E_-=C_-^{(1)}(1-\mu^2)^\lambda w_3+C_-^{(2)}(1-\mu^2)^\lambda w_4$. According to Appendix A, the asymptotic behaviors for $x\rightarrow\pm\infty$ or $y\approx1$ are that $w_3\approx1$ and $w_4\approx(1-y)^{-2\lambda}$.

Therefore, for $x\rightarrow+\infty$,
\begin{equation}
E_+\approx(1-\mu^2)^\lambda\approx e^{ix\rho\sqrt{a^2-b^2}}e^{-\delta\sqrt{a^2-b^2}}, \label{out}
\end{equation}
which is an outgoing wave; for $x\rightarrow-\infty$,
\begin{equation}
E_-\approx C_-^{(1)} e^{-ix\rho\sqrt{a^2-b^2}}e^{+\delta\sqrt{a^2-b^2}}+C_-^{(2)} e^{ix\rho\sqrt{a^2-b^2}}e^{-\delta\sqrt{a^2-b^2}}, \label{in-reflect}
\end{equation}
where $C_-^{(1,2)}$ term represents incident and reflected wave, respectively, since the energy flux and wave vector point in opposite directions in this double negative region.

From Eq.(\ref{out},\ref{in-reflect}) above, we can directly read out the reflection $\mathcal{R}$ and transmission $\mathcal{T}$ coefficients,
\begin{eqnarray}
\mathcal{R}&=&\frac{C_-^{(2)}}{C_-^{(1)}}e^{-2\delta\sqrt{a^2-b^2}}\nonumber\\
\mathcal{T}&=&\frac{1}{C_-^{(1)}}e^{-2\delta\sqrt{a^2-b^2}}
\end{eqnarray}

\section{Maximum Total absorption}
Before we investigate the total absorption in the lossless limit $\delta\rightarrow0$, we first prove the identity $|\mathcal{R}|+|\mathcal{T}|=1$, or equivalently $|C_-^{(1)}|=1+|C_-^{(2)}|$. We can easily see that $|C_-^{(1)}|=C_-^{(1)}>1$, since $\alpha$ and $\beta$ are purely imaginary. Then we calculate $|C_-^{(2)}|$.
\begin{equation}
|C_-^{(2)}|^2=C_-^{(2)}C_-^{(2)*}=C_-^{(2)}(\alpha,\beta)C_-^{(2)}(-\alpha,-\beta)=-\frac{4\sin^2\pi \beta\sin^2\pi\alpha}{\sin^2\pi(\alpha+\beta)}.
\end{equation}
Then follows the identity $|C_-^{(1)}|=1+|C_-^{(2)}|$.

Given the above identity, the total absorption $\mathcal{A}=1-|\mathcal{R}|^2-|\mathcal{T}|^2$ reaches its maximum 1/2 under the condition $|\mathcal{R}|=|\mathcal{T}|=\mathcal{T}=1/2$. Then we finally arrive at the incident angle for maximum absorption,
\begin{equation}
\cos\theta=\frac{\ln\cosh \pi\omega/\rho}{\pi\omega/\rho}.
\end{equation}

\section{$\mathcal{R}$ and $\mathcal{T}$ in several limiting cases:}
We only consider the limit $\delta\rightarrow0$ in this section.

For a nonzero angle $\theta\neq0$ and high frequency $\omega\rightarrow+\infty$, it can be easily proved that $|C_-^{(2)}|\rightarrow+\infty$. Therefore $\lim_{\omega\rightarrow+\infty}|\mathcal{T}|=0$, and $\lim_{\omega\rightarrow+\infty}|\mathcal{R}|=1$.

For a nonzero angle $\theta\neq0$ and low frequency $\omega\rightarrow0$, we have $\alpha\rightarrow0$ and $\beta\rightarrow0$, followed by $|C_-^{(2)}|=0$, $|C_-^{(1)}|=1$ and $\lim_{\omega\rightarrow0}|\mathcal{T}|=1$, $\lim_{\omega\rightarrow0}|\mathcal{R}|=0$.

For zero angle $\theta=0$ and a nonzero $\omega$, we can easily get $\lim_{\theta\rightarrow0}|\mathcal{T}|=1$, $\lim_{\theta\rightarrow0}|\mathcal{R}|=0$.

For large angle near $\pi/2$ and a nonzero $\omega$, we can prove that $|C_-^{(2)}|\rightarrow+\infty$. Then follows $\lim_{\theta\rightarrow\pi/2}|\mathcal{T}|=0$, and $\lim_{\theta\rightarrow\pi/2}|\mathcal{R}|=1$.

\section{Effects on perfect imaging}
In Ref.\onlinecite{Pendry2000}, it is proposed that a slab (perfect lens) with $\epsilon=\mu=-1$ can overcome the diffraction limit. The mechanism is that the perfect lens can restore the information in the evanescent waves.

In the sharp transition limit, $\rho\rightarrow+\infty$, the transition materials approaches a half-infinite perfect-lens slab. If we also assume lossless limit $\delta\rightarrow0$, we can obtain a perfect image of an object in another side of the interface since the transmission for all transverse wave vectors are unity.

However, as can be seen from previous sections, the transmission is not unity even for the propagating wave if the transition is smooth, i.e., $\rho$ is finite. If we set a monochromatic object (input) in the plane $x=-a$, with its transverse amplitude profile $E(x=-a;z)=\frac{\Delta}{\pi}\frac{1}{z^2+\Delta^2}$ and Fourier components $E(x=-a;k)=e^{-\Delta|k|}$, we will obtain the field amplitudes in the image plane $x=a$,
\begin{equation}
E(x=a;z)=\frac{1}{2\pi}\int E(x=-a;k)T(k)e^{iky}dk,
\end{equation}
where, according to the results in Sec.\ref{SecRT},
\begin{equation}
T(k)=\frac{\sin2\pi\lambda}{e^{-2i\pi\lambda}-\cosh\pi\omega/\rho},
\end{equation}
with $\lambda=-\sqrt{k^2-\omega^2}/2\rho$ being the principal value of the square root. We have also assumed that $\delta\rightarrow0$.

Because $T$ is a periodic function of $\lambda$, we can obtain the discrete Fourier expansion of $T$,
\begin{equation}
T=\sum_{-\infty}^{+\infty}a_n \exp^{2i\pi n\lambda},
\end{equation}
with $a_n=\frac{i}{2}(\chi^{-2}-1)\chi^{-n}$ for $n=1,2,\cdots$; $a_0=\frac{i}{2}\chi^{-2}$; $a_{-1}=\frac{i}{2}\chi^{-1}$; $a_n=0$ for $n=-2,-3,\cdots$. Here $\chi=\cosh(\pi\omega/\rho)$.

A remarkable feature of each Fourier term $\exp(i\pi n\sqrt{(k^2-\omega^2)/\rho})$ is that for large $k$, it approaches a periodic function and thus contributes a resonance when anti-Fourier-transformed. Let us approximate $\exp{i\pi n\sqrt{(k^2-\omega^2)/\rho}}\sim\exp{i\pi n|k|}$ for large $k$. Then the resonances in the real $z$ axis can easily be obtained after anti-Fourier transformation.
\begin{equation}
E(x=a;z)\sim\sum_{n=-\infty}^{n=+\infty}\frac{a_n}{\pi}\frac{n\pi/\rho-i\Delta}{z^2-(n\pi/\rho-i\Delta)^2}.\label{multi}
\end{equation}

The above result indicates that the image plane contains multiple images of the object, a Lorentzian spot. The widths of the all images remain the same as that of the object. The amplitude $a_n$ of high order images exponentially decays with a characteristic order $n_c=\frac{1}{\ln\cosh(\pi\omega/\rho)}$.
\section{Summary}
Analytical solutions are obtained for smooth transition material with ``tanh'' profile. The field is expressed in terms of hypergeometric functions. We also obtain the closed-form analytical expressions for the reflection and transmission coefficients. The total absorption for this transition material is none zero even in the lossless limit. The maximum of the total absorption, $\frac{1}{2}$, is also analytically located. The asymptotic behaviors for several limiting cases have also been considered. Our analytical results can shed light on the smooth transition effects on imaging: emerging of multiple images.

\section*{acknowledgement}
We thank Meng Xiao, Chiu Yin Tsang, Edmund Chiang for helpful discussions.

\appendix
\section{Kummer solutions for hypergeometric equations}
For convenience, we display the knowledge about hypergeometric equations used in this study. One can also refer to \verb|http://dlmf.nist.gov/15.10|. The hypergeometric equation is
\begin{equation}
z(1-z)\frac{d^{2}w}{dz^{2}}+[\gamma-(1+\alpha+\beta)z]\frac{dw}{dz}-\alpha\beta w=0.
\end{equation}

When none of $\gamma$, $\gamma-\alpha-\beta$, $\alpha-\beta$ is an integer, the two basic solutions near $0<|z|<1$ are,
\begin{eqnarray}
w_1(z)&=&~_2F_1(\alpha,\beta;\gamma;z),\nonumber\\
w_2(z)&=&z^{1-\gamma}~_2F_1(\alpha-\gamma+1,\beta-\gamma+1;2-\gamma;z);\label{w1w2}
\end{eqnarray}
the two basic solutions for $0<|z-1|<1$ are,
\begin{eqnarray}
w_3(z)&=&~_2F_1(\alpha,\beta;\alpha+\beta-\gamma+1;1-z),\nonumber\\
w_4(z)&=&(1-z)^{\gamma-\alpha-\beta}~_2F_1(\gamma-\alpha,\gamma-\beta;\gamma-\alpha-\beta+1;1-z),
\end{eqnarray}
where
\begin{equation}
~_2F_1(\alpha,\beta;\gamma;z)=\sum_{n=0}^{+\infty}\frac{1}{n!}\frac{\Gamma(\alpha+n)}{\Gamma(\alpha)}\frac{\Gamma(\beta+n)}{\Gamma(\beta)}
\frac{\Gamma(\gamma)}{\Gamma(\gamma+n)}z^n,
\end{equation}
and definitions of these solutions on the whole complex plane are not displayed here.

They are connected by Kummer connection formulas in common converging regions, which include
\begin{equation}
w_{3}=\frac{\Gamma(1-\gamma)\Gamma(\alpha+\beta-\gamma+1)}{\Gamma(\alpha-\gamma+1)\Gamma(\beta-\gamma+1)}w_{1}+
\frac{\Gamma(\gamma-1)\Gamma(\alpha+\beta-\gamma+1)}{\Gamma(\alpha)\Gamma(\beta)}w_{2};
\end{equation}
\begin{equation}
w_{2}=\frac{\Gamma(2-\gamma)\Gamma(\gamma-\alpha-\beta)}{\Gamma(1-\alpha)\Gamma(1-\beta)}w_{3}+
\frac{\Gamma(2-\gamma)\Gamma(\alpha+\beta-\gamma)}{\Gamma(\alpha-\gamma+1)\Gamma(\beta-\gamma+1)}w_{4}
\end{equation}

\section{Logarithmic solution near $z=0$ for $\gamma\rightarrow0$}
To prove Eq.(\ref{f}), we only need to guarantee that its numerator $\frac{w_{1}}{\Gamma(\gamma)}-\alpha\beta w_{2}$ approaches zero when $\gamma\rightarrow0$.

According to Eq.(\ref{w1w2}), when $\gamma\rightarrow0$,
\begin{eqnarray}
w_1&\approx&1+\Gamma(\gamma)\alpha\beta z\sum_{n=0}^{+\infty}\frac{1}{n!}\frac{\Gamma(\alpha+n+1)}{\Gamma(\alpha+1)}\frac{\Gamma(\beta+n+1)}{\Gamma(\beta+1)}
\frac{\Gamma(\gamma+2)}{\Gamma(\gamma+n+2)}z^n\nonumber\\
&\approx&1+\Gamma(\gamma)\alpha\beta z~_2F_1(\alpha+1,\beta+1;2;z);
\end{eqnarray}
and
\begin{eqnarray}
w_2&\approx&z~_2F_1(\alpha+1,\beta+1;2;z);
\end{eqnarray}
Then the Eq.(\ref{f}) can be understood.

%

\end{document}